\documentclass[]{spie}  

\usepackage{amsmath,amsfonts,amssymb}
\usepackage{graphicx}
\usepackage[colorlinks=true, allcolors=blue]{hyperref}\usepackage{xspace}
\usepackage{upgreek}
\usepackage{amsmath,amsfonts,amssymb}
\usepackage{graphicx}
\usepackage{color, colortbl} 
\usepackage{booktabs} 
\usepackage{multirow} 
\usepackage[inline]{enumitem} 
\usepackage{wrapfig}

\usepackage{wasysym}
\usepackage{physics} 
\usepackage{textcomp,gensymb} 
\usepackage{subcaption} 
\usepackage{siunitx}

\newcommand{\um}{$\upmu$m\xspace}	

\def\arcmin{\hbox{$^\prime$}\xspace} 

\def\procspie{\ref@jnl{Proc.~SPIE}}   

\title{Characterization, deployment, and in-flight performance of the BLAST-TNG cryogenic receiver}

\author[a]{Ian Lowe}
\author[b]{Peter A. R. Ade}
\author[c]{Peter C. Ashton}
\author[d]{Jason E. Austermann}
\author[e]{Gabriele Coppi}
\author[f]{Erin G. Cox}
\author[a]{Mark J. Devlin}
\author[d]{Bradley J. Dober}
\author[e]{Valentina Fanfani}
\author[g]{Laura M. Fissel}
\author[h]{Nicholas Galitzki}
\author[d]{Jiansong Gao}
\author[i]{Samuel Gordon}
\author[i]{Christopher E. Groppi}
\author[d]{Gene C. Hilton}
\author[d]{Johannes Hubmayr}
\author[a]{Jeffrey Klein}
\author[j]{Dale Li}
\author[k]{Nathan P. Lourie}
\author[i]{Hamdi Mani}
\author[i]{Philip Mauskopf}
\author[d]{Christopher McKenney}
\author[e]{Federico Nati}
\author[f]{Giles Novak}
\author[b]{Giampaolo Pisano}
\author[l]{L. Javier Romualdez}
\author[n]{Juan D. Soler}
\author[i]{Adrian Sinclair}
\author[b]{Carole Tucker}
\author[d]{Joel Ullom}
\author[d]{Michael Vissers}
\author[m]{Caleb Wheeler}
\author[f]{Paul A. Williams}

\affil[a]{University of Pennsylvania, 209 South 33rd Street, Philadelphia, PA 19104, USA}
\affil[b]{Cardiff University, The Parade, Cardiff CF24 3AA, United Kingdom}
\affil[c]{Lawrence Berkeley National Laboratory, 1 Cyclotron Rd, Berkeley, CA 94720}
\affil[d]{NIST-Boulder, 325 Broadway, Boulder, CO 80305}
\affil[e]{University of Milano-Bicocca, Piazza della Scienza 3, 20126 Milano (MI), Italy}
\affil[f]{Northwestern University, 1800 Sherman Ave, Evanston, IL 60201}
\affil[g]{Queen's University, 99 University Ave, Kingston, ON K7L 3N6, Canada}
\affil[h]{University of California San Diego, 9500 Gilman Dr, La Jolla, CA 92093}
\affil[i]{Arizona State University, 550 E Tyler Drive, Tempe, AZ 85287}
\affil[j]{SLAC National Accelerator Laboratory, 2575 Sand Hill Rd, Menlo Park, CA 94025}
\affil[k]{MIT Kavli Institute for Astrophysics and Space Research, 70 Vassar St, Cambridge, MA 02139}
\affil[l]{Princeton University, Jadwin Hall, Washington Road, Princeton, NJ 08544}
\affil[m]{Underground Instruments, 224 Claremont Ave. San Antonio TX 78209}
\affil[n]{Max-Planck-Institut für Astronomie, Königstuhl 17
69117 Heidelberg, Germany}

\authorinfo{Further author information: (Send correspondence to Ian Lowe)\\Ian Lowe: E-mail: ianlowe@sas.upenn.edu, Telephone: 1 484 919 6228}

\begin{document} 
\maketitle

\begin{abstract}
\paragraph{}

The Next Generation Balloon-borne Large Aperture Submillimeter Telescope (BLAST-TNG) is a submillimeter polarimeter designed to map interstellar dust and galactic foregrounds at 250, 350, and 500 microns during a 24-day Antarctic flight. The BLAST-TNG detector arrays are comprised of 918, 469, and 272 MKID pixels, respectively. The pixels are formed from two orthogonally oriented, crossed, linear-polarization sensitive MKID antennae. The arrays are cooled to sub 300~mK temperatures and stabilized via a closed cycle $^3$He sorption fridge in combination with a $^4$He vacuum pot. The detectors are read out through a combination of the second-generation Reconfigurable Open Architecture Computing Hardware (ROACH2) and custom RF electronics designed for BLAST-TNG. The firmware and software designed to readout and characterize these detectors was built from scratch by the BLAST team around these detectors, and has been adapted for use by other MKID instruments such as TolTEC and OLIMPO\cite{Paiella_2019}. We present an overview of these systems as well as in-depth methodology of the ground-based characterization and the measured in-flight performance.
\end{abstract}
\section{Introduction}
\label{section:intro}
\paragraph{}

The Next Generation Balloon-Borne Large Aperture Submillimeter Telescope (BLAST-TNG) is a unique instrument for characterizing the polarized submillimeter sky at high angular resolution. BLAST-TNG is designed to study the role of magnetic fields in shaping the structure and evolution of the interstellar medium (ISM) and the role of magnetic fields in regulating star formation by mapping polarized dust emission. BLAST-TNG is also designed to have a high range of angular resolution and sensitivity to complement  current and planned submillimeter observatories like \textit{Planck} \cite{planck_XIX}, ALMA \cite{alma}, \textit{SOFIA} \cite{sofia}, POL-2 on the James Clark Maxwell Telescope \cite{pol2_jcmt}, and CCAT-Prime \cite{ccat_prime_parshley}. 

BLAST-TNG features three microwave kinetic inductance detector (MKID) arrays operating over 30\% bandwidths centered at 250, 350, and 500 \um. In the three bands, these highly-multiplexed, high-sensitivity arrays include 918, 469, and 272 dual-polarization pixels respectively, for a total of more than 3,000 detectors. The detectors are coupled to a cassegrain telescope through a cold reimaging optics system resulting in a diffraction-limited instrument with a resolution of 30, 41 and 59 arcsec. Simultaneous observations are achieved through the use of two dichroic beam splitters. The arrays are cooled to $\sim$275~mK in a liquid-helium-cooled cryogenic receiver, with a 250~L reservoir which can operate for \textgreater 24 days. BLAST-TNG was launched on January 6th, 2020 and made observations during an 18 hour flight from McMurdo Station in Antarctica as part of NASA's long-duration-balloon (LDB) program. Despite the short flight, the team was able to get valuable data about the performance of the detectors and other subsystems that would prove their functionality on a balloon platform. Much of the instrument (including the receiver) was destroyed due to a rough impact and subsequent drag upon landing, however the primary mirror and many subsystems (such as the readout and control electronics, motors, and hard drives) were successfully recovered during an intense recovery campaign.

The BLAST-TNG observations were designed to fulfill several scientific goals, including to: 
\begin{enumerate*}[label=(\arabic*)] 
\item obtain a sample of deep maps of star-forming regions with sizes ranging from 0.25 to 20 square degrees to better understand how magnetic fields influence the formation and evolution of molecular clouds;
\item apply statistical methods to quantify the observed magnetic field morphology. By measuring the magnetic field, its dispersion, and their correlations with other physical parameters of the ISM, constraints can be put on field strength, the turbulent power spectrum, and models of cloud formation\cite{houde_2009_turbulence, soler_hennebelle, soler_sims_2013, jow_hro, fissel_blastpol};
\item map magnetic fields in a large sample of dense prestellar and protostellar cores, and show how the core magnetic field relates to the larger scale cloud fields. 
\item map diffuse interstellar dust to constrain models of dust composition and alignment mechanisms through measurements of changes in polarization as a function of frequency \cite{draine_fraisse_2009,draine_hensley};
\item characterize the (polarized) diffuse dust emission on small angular scales \item serve as a true ``observatory'' platform on which science groups outside of the BLAST-TNG collaboration can propose observations, up to 25$\%$ of the observation time. \end{enumerate*}

This paper describes the design, operation, and in-flight performance of the critical components of the BLAST-TNG payload. In Section \ref{section:characterization}, we describe the characterization and testing prescription for the cryogenic receiver, Section \ref{section:deployment} covers the deployment and further testing of the instrument, and Section \ref{section:inflight} reviews the in-flight performance of the receiver. Finally, in Section \ref{section:concs}, we conclude about the performances of the subsystems as well as their future utility.

\section{Characterization of the BLAST-TNG receiver}
\label{section:characterization}
\begin{wrapfigure}{l}{0.5\textwidth}
    \centering
    \includegraphics[scale=1]{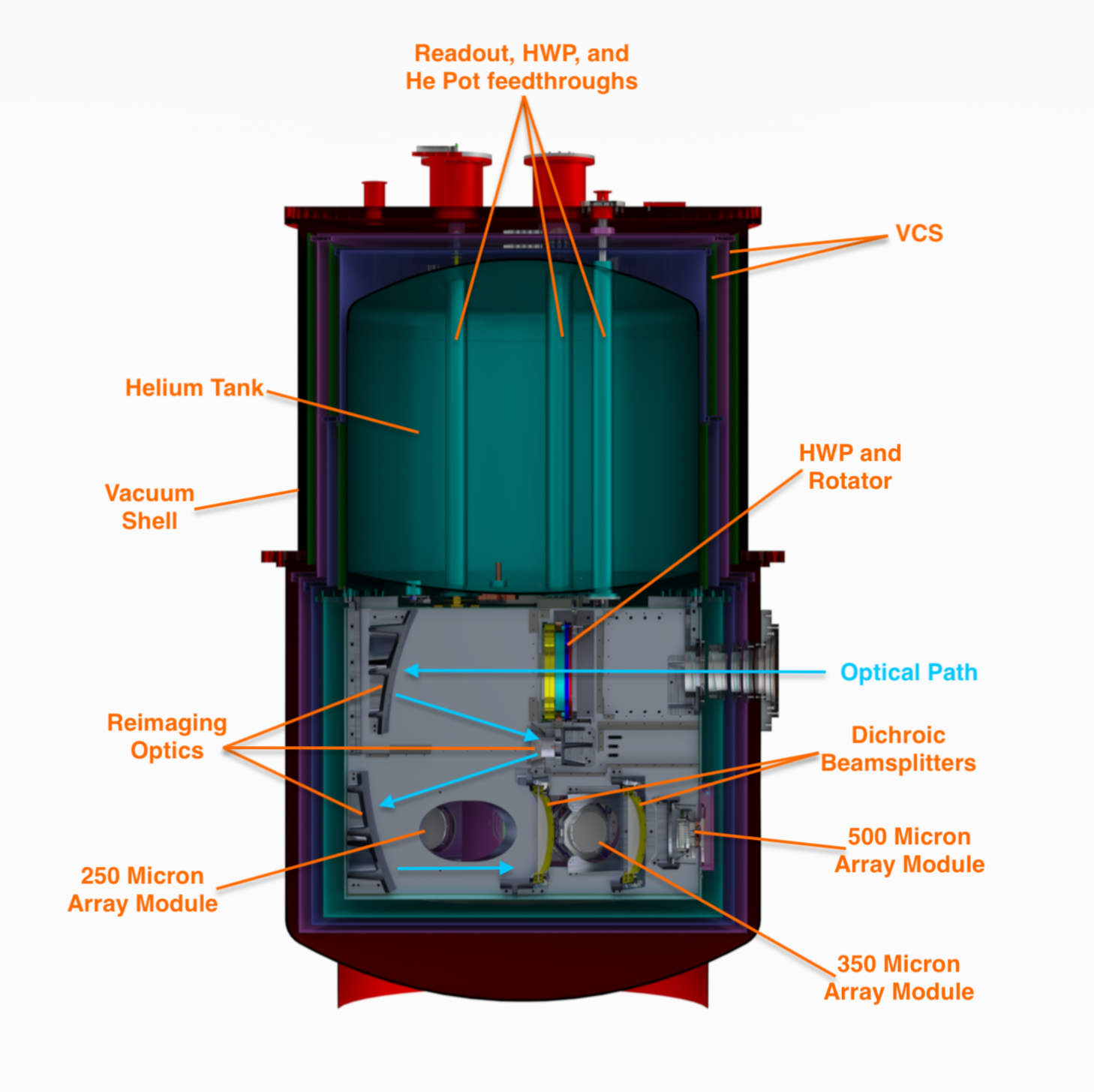}
    \caption{Cutaway model of the BLAST-TNG cryogenic receiver highlighting the optics box elements and nested structure of the receiver. Refrigeration and control systems on the 4~K plate are not visible from this side.}
    \label{fig:cryostat}
\end{wrapfigure}

\paragraph{}
In this section, we discuss the characterization and integration of the subsystems of the BLAST-TNG cryogenic receiver (Figure \ref{fig:cryostat}). We detail the testing for the cryogenics and detectors leading up to the deployment to Palestine and Antarctica. We also discuss the baseline configuration and design of the receiver and readout in order to familiarize the reader. Additional information regarding the instrument and receiver can be found in papers on BLAST-TNG\cite{tyr_blasttng_spie, Nate_det_spie}.
\subsection{Receiver and Cryogenics}
\paragraph{}
The BLAST-TNG receiver is a 250~L liquid helium-based cryostat, designed to provide a $<300$~mK enviroment for the MKIDs for a $>24$ day flight. The receiver is based off of the previous design\cite{enzo_blast}, though larger and without a liquid nitrogen tank. The vacuum shell surrounds a series of concentric aluminum shields offset and connected by G10 cylinders to minimize parasitic loading. Each layer is covered by superinsulation (RUAG Space) which serves to minimize optical loading between the layers. The cryostat houses a pair of nested vapor-cooled shields (VCS) which equilibrate at roughly 140 and 40~K through a pair of copper heat exchangers integrated into the vent tube. The innermost shell sits at 4~K and comprises the helium tank as well as the cold plate on which the optics box, refrigeration systems, and cryogenic electronics sit. The optical path into the receiver consists of a series of reflecting IR-blocking filters, three low-pass filters which serve to step down the high-frequency edge of the band, a halfwave plate which rotates the polarization of the incident light, reimaging mirrors and a Lyot stop to minimize side lobes, dichroic beam splitters, and a final band-definition filter caps each array. The filters were all designed by the QMC Instruments group and provide a significant reduction in the optical loading such that the final stage at 300~mK receives minimal out of band radiation and heating.

The BLAST-TNG refrigeration system was designed to provide high duty-cycle, quick cycling, and sub-300~mK temperatures at the arrays. The base of the refrigeration system is the 4~K cold plate and helium bath that provides a stable base temperature with an enormous capacity to absorb power transferred into it without large changes in temperature. The middle stage runs between 1.09-1.15~K during normal operating conditions and is comprised of a 200~ml volume of liquid helium which is valved off from the tank on one end and exposed to 600~$\mu$Bar on the other end. The final stage is the $^3$He sorption fridge which is a sealed and recyclable volume of helium that utilizes activated charcoal fins on one end to adsorb the vapor and provide a near-vacuum above the liquid helium. This system provides an operating temperature of 265~mK at the evaporating still or 274~mK at the arrays. The sorption fridge is coupled to the pumped pot at the condensing ring which allows for recycling when the charcoal is heated. The high duty cycle and quick cycling are simultaneously achieved through the large cooling capacity of the pumped pot, which is able to provide several days of 1K temperatures as well as quickly remove heat generated during the recycling process to cool the sorption fridge back to base temperature. A typical cycle takes approximately 1.5 hours from start to finish and is only required every 3 days for an effective duty cycle of 98$\%$.
\subsection{Readout and Detectors}
\subsubsection{Detectors}
\paragraph{}

\begin{figure}
    \centering
    \includegraphics[scale=1]{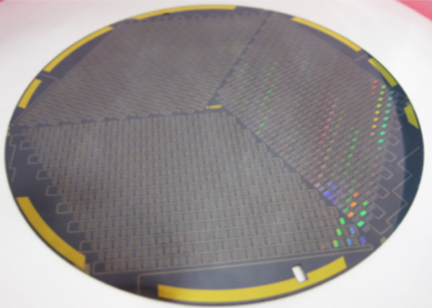}
    \includegraphics[scale=0.436]{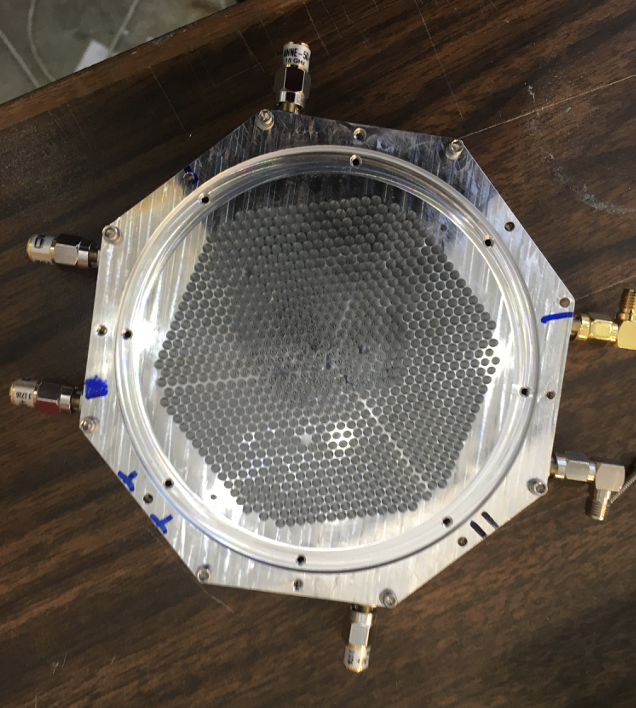}
    \caption{(left) Bare BLAST-TNG 250~$\mu m$ array. (right) Mounted BLAST-TNG 250~$\mu m$ array with feedhorn block over the top, prior to the installation of the final filter.}
    \label{fig:array}
\end{figure}

The detector arrays (Figure \ref{fig:array}) consist of hundreds of MKIDs coupled to a microwave feedline to be read out over a single coaxial cable. An MKID acts as a high-Q resonant LC circuit with a characteristic frequency, which is then grounded, forming a notch filter. During fabrication, the capacitance of each detector is varied slightly, which creates a comb of resonances across a frequency band designed to match the readout capabilities. Light incident on the optical element of the detectors causes Cooper pairs to break, changing the inductance of the detector and shifting the resonance until they reform. This results in an power dependent frequency shift, while maintaining the shape of the resonance for small incident powers. We take advantage of this, and knowledge of the frequency response to different loading to turn a shift into measurements of the sky power incident on the detectors.

\subsubsection{Readout}
\paragraph{}

BLAST-TNG takes advantage of highly frequency multiplexed polarized MKID detector arrays to maximize detector count while minimizing the number of coaxial lines used for reading out the arrays. The readout system is based off of ROACH2, designed by the Collaboration for Astronomy Signal Processing and Electronics Research\cite{werthimer2011} (CASPER) which works with a flight computer to execute the readout instructions. The BLAST-TNG collaboration developed the firmware to generate the readout functionality as well as the python-based software KIDPy which implements the instructions for reading the detectors out. A detailed description of the BLAST readout hardware, firmware design, and implementation can be found in our detector papers\cite{sam_roaches,sam_thesis}.

\begin{wrapfigure}{r}{0.5\textwidth}
    \centering
    \includegraphics[scale=.48]{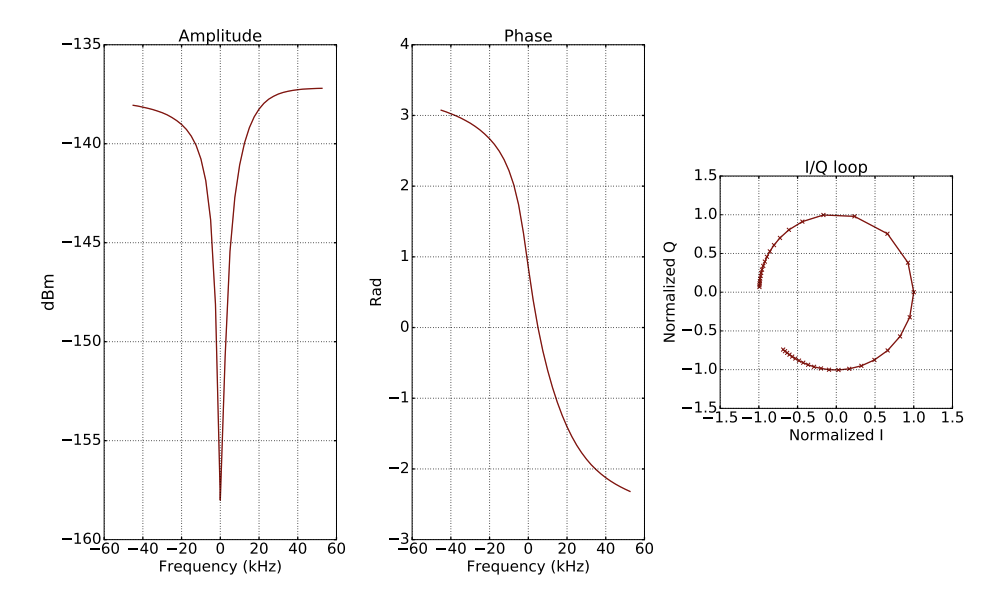}
    \caption{Target sweep products showing the mapping of an individual resonator from early detector and readout testing\cite{sam_roaches}.}
    \label{fig:targ}
\end{wrapfigure}

The readout depends on a KID-finding algorithm that determines the locations of the KIDs, and maps them to initiate functionality before sending the appropriate tone comb into the detectors to read them out. The first step in this algorithm is to make a measurement of the array in frequency space with the ROACH. This is done by creating a comb of 1000 tones across the 512~MHz system bandwidth and using the local oscillator to sweep the space in between each tone at a resolution of 2.5~KHz. This ``VNA sweep" mode creates a rough map of the S21 response in frequency space, which is then processed using peak finding algorithms to determine and save the locations of the resonances. These saved tones are what will be continuously fed into the detectors as the readout comb. We then utilize the ``Target sweep" mode, which sweeps 100~KHz of bandwidth centered on each resonance location and saves the corresponding individual resonator maps for the final stage in detector preparation. With these resonator maps, we measure (Figure \ref{fig:targ}) the shape in phase and magnitude response of the detectors, specifically in the linear regime where a frequency shift produces a linear change in the signal received by the ROACH. By sending a constant comb of tones into the arrays and analyzing the varying response power at each frequency, we can convert the change in received signal from the baseline to a shift in resonant frequency.

While using these detectors it is important to ensure that the probe tones remain properly located with respect to the resonators. This is because changes in static loading or temperature cause them to shift in frequency and potentially go ``off-resonance" if the changes are drastic. To ensure that our frequency comb remains centered on the correct locations, we utilize a calibration lamp located in the Lyot stop of the reimaging optics designed to produce a repeatable signal. Flashing the lamp after the target sweep and at specific intervals during operations allows us to compare the response of the detectors and determine, based on the relative response, if reinitiating the KID-finding algorithm is required.

\subsection{Testing methods}
\label{sec:testing}
\paragraph{}

\begin{figure}
    \centering
    \includegraphics[scale=.45]{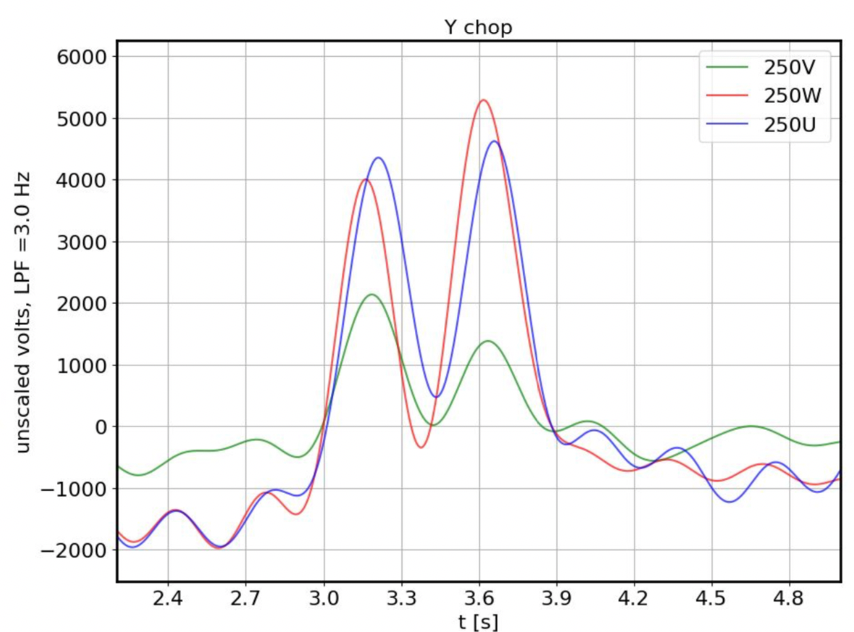}
    \caption{Proof of concept measurement of the response of detectors as a hot source scanned in the vertical direction in front of the cryostat window showing the bimodal response expected from the Lyot stop aperture.}
    \label{fig:lyot}
\end{figure}

To initially characterize the receiver, we used several distinct testing methods, each aimed at providing an understanding of a different component of the detector functionality. The first and most widely used testing device was the cal lamp mentioned in the readout discussion. This method of testing was mainly used in the initial characterization for demonstration of the readout functionality, seeing the oscillations in the time stream, as well as readout optimization, which involved maximizing the signal to noise ratio (SNR) of the cal lamp pulses by modifying software settings within the readout. Early testing also involved the use of liquid nitrogen sources, which provided a much lower background loading level when placed in the optical path to test detector properties at different loading. A crucial piece of testing equipment was the heated beam-filling source that provided a low frequency chop between a rotating room temperature black plate and a resistive plate warmed to 10-25~\degree C above room temperature. By filling the beams with this source and chopping between two known temperatures we were able to perform initial responsivity measurements on the detectors in the receiver. Finally, we used a high temperature, small-aperture ceramic heating element for  a proof-of-concept measurement so we could see the Lyot stop (Figure \ref{fig:lyot}) as we scanned it across individual beams. 

\section{Deployment}
\label{section:deployment}
\paragraph{}

BLAST-TNG was deployed to the Columbia Scientific Balloon Facility (CSBF) for final integration and compatibility testing in July 2018. During this time period, the cryogenic receiver, detector, and readout systems all underwent rigorous testing and system upgrades before the launch campaign in Antarctica in November 2018. During the 2018/2019 Antarctic campaign, it was discovered that one of the readout chains had been broken in shipping after the experiment was cooled down. This could not be fixed due to the time constraints of the season and the hazard fixing it could pose to the rest of the readout. Following the cancellation of the season due to the break up of the polar vortex, the readout chain was earmarked for a simple redesign and installation while the experiment was left on the ice. In this section, we detail the upgrades to, tests done with, and performance of the cryogenic receiver and readout during the Palestine, 2018/2019 Antarctic, and 2019/2020 Antarctic campaigns. 
\subsection{Palestine}
\paragraph{}

In Palestine, the detectors and readout underwent a significant amount of testing to prove that they would be ready for flight, as well as continued upgrades to the hardware to further improve performance. Over the course of the deployment, we took new noise measurements to understand the improved sensitivity of the detectors, performed a bandpass measurement on the arrays using a Fourier transform spectrometer (FTS), made the first maps of the detector arrays in the cryostat (350~$\mu m$ perviously mapped at NIST), upgraded the ROACH systems with new low noise amplifiers and filters, and modified the cryogenic RF pathway to optimize the attenuation of the system. In addition to these tests, we continued to use the standard cal lamp, liquid nitrogen source, hot plate, and ceramic IR sources as litmus tests of detector functionality. However, these new tests provided further insight into the properties of the instrument and ensured that we would have sufficient understanding to be flight ready. Finally, the cryogenics were shown to have improved significantly over earlier cooldowns, due to the blackening of the optics box to absorb stray hot light and the vastly improved performance of the new 4K charcoal adsorptive element.

\begin{figure}
    \centering
    \includegraphics[scale=0.35]{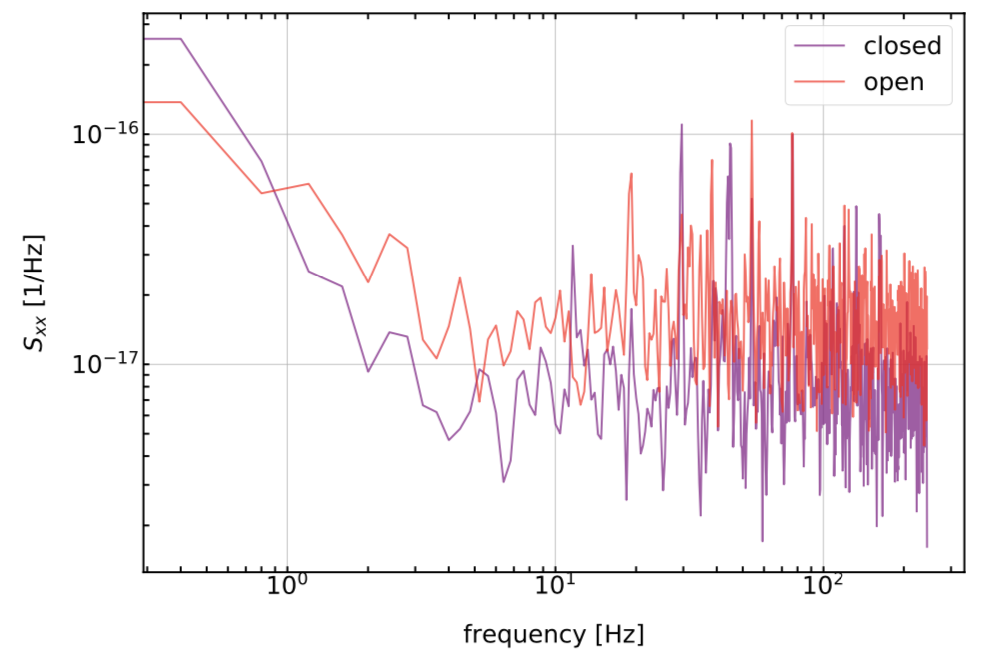}
    \includegraphics[scale=0.29]{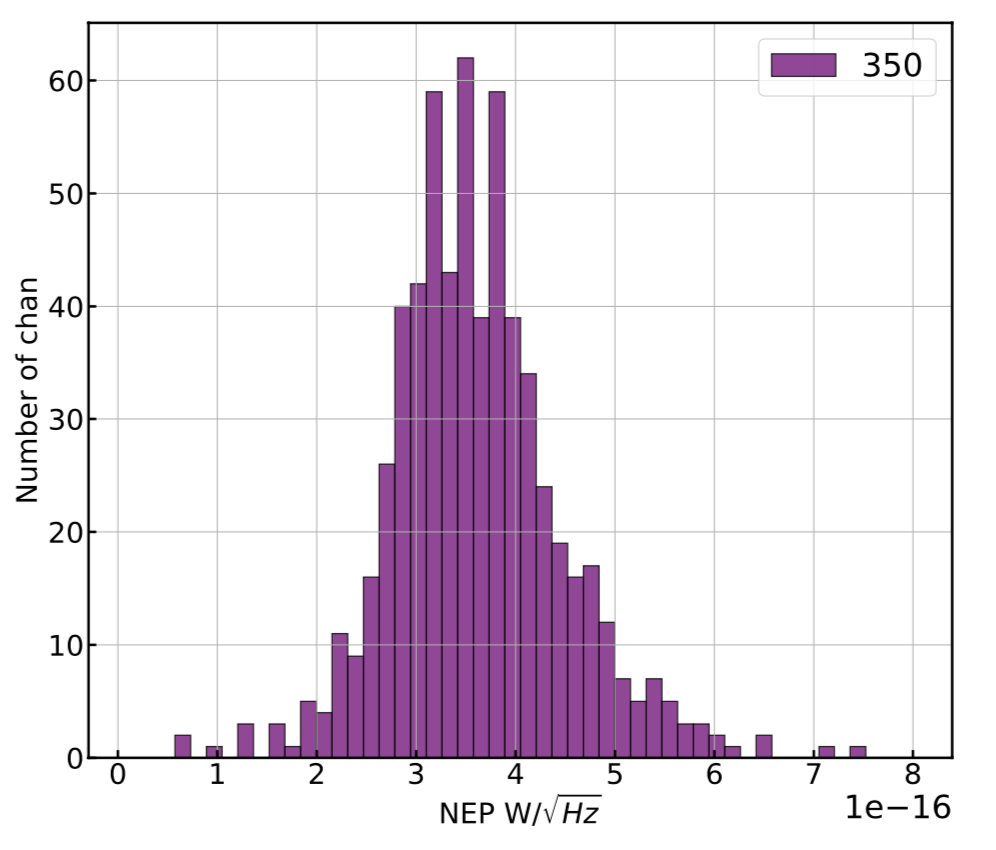}
    \caption{(right) Two timestreams processed into PSDs showing the higher fractional frequency noise seen in the detectors when looking into a 300~K room versus a parabolic reflective plate. (left) Histogram of detector electrical NEPs from measurements in Palestine. These measurements give us a crucial understanding of the noise level that our electronics contribute to the experiment.}
    \label{fig:noisepal}
\end{figure}

The characterization of the detectors started with new noise measurements designed to compare the previous and planned sensitivity of the receiver and readout system to the newly upgraded version. To take these noise
\begin{wrapfigure}{l}{0.5\textwidth}
    \centering
    \includegraphics[scale=0.3]{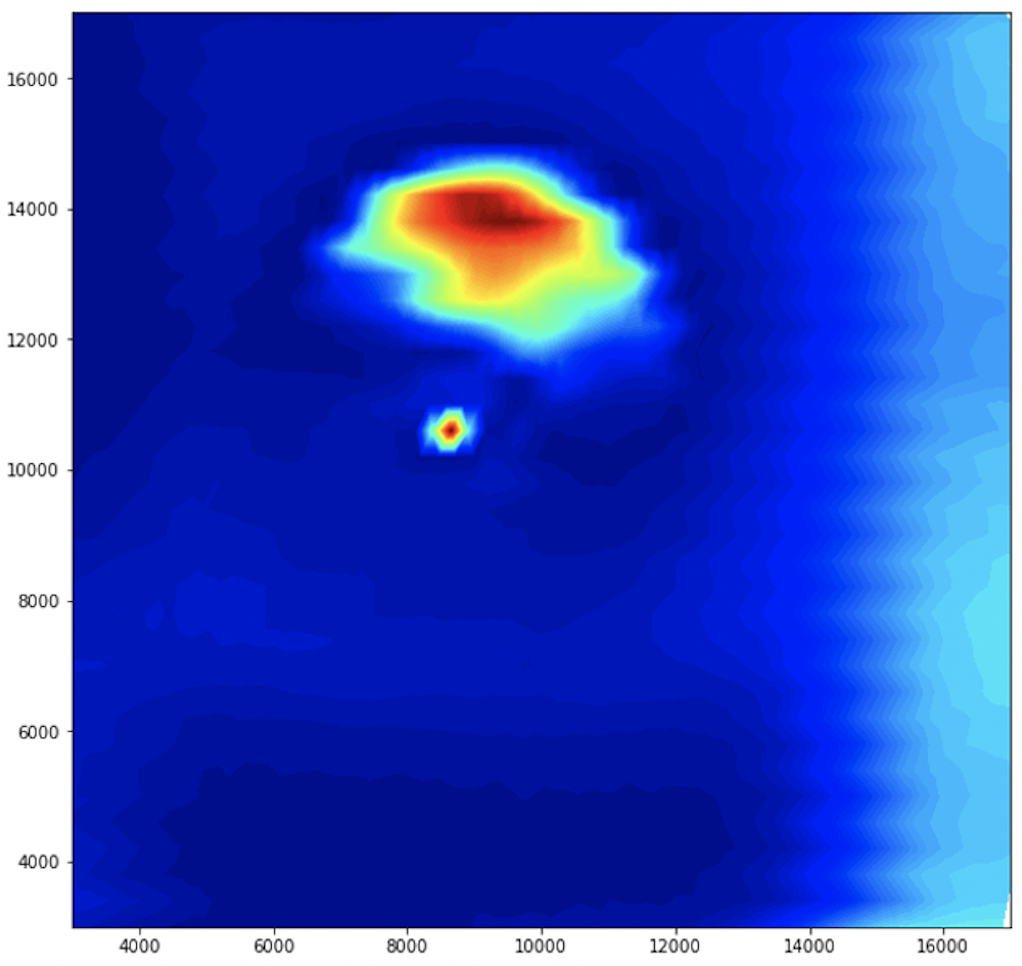}
    \includegraphics[scale=0.5]{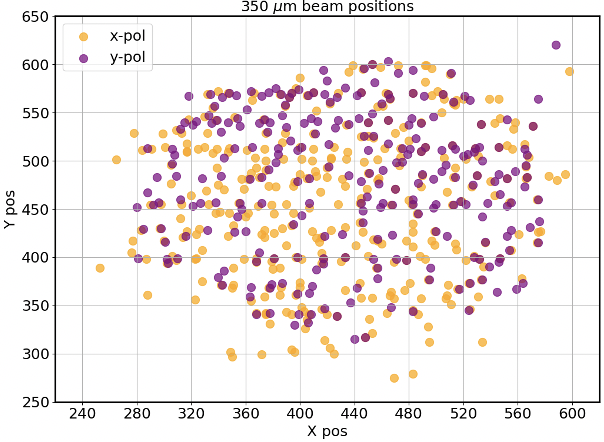}
    \caption{(top) Single-detector map of a scan. The small blob is the source response and the larger an unknown contamination. (bottom) Best fit array map with polarization from the test.\cite{sam_thesis}}
    \label{fig:arraymap}
\end{wrapfigure} data, we record timestreams from each of the detectors over approximately 10 seconds under differing conditions. These conditions may be looking at a cold source, looking into the room, looking at a hot source,
or looking at a reflective metal cover. While all different loading conditions, these each provide a quasi-static thermal environment to understand the noise level under different loading conditions. The timestreams are processed into a noise level (NEP per detector) and into power spectral density (PSD) graphs to elucidate the frequency dependence (Figure \ref{fig:noisepal}).

Our next set of tests was the preliminary mapping of the arrays. The 350 array had been mapped using an LED attachment at NIST, however the two other arrays had not. To map the arrays we attached a thermal source to a programmable XY-stage which was mounted in front of the receiver window with a polarizing grid. The stage was then scanned in a square pattern to cover the entire field of view before rotating the grid to admit the other polarization of light and repeating. We combined the position information with the timestreams to create maps from each of the detectors and translated this information into array maps. The setup resulted in some false positives and large sources of unknown emission as seen in Figure \ref{fig:arraymap} and added some dispersion to our array map.

\begin{figure}
    \centering
    \includegraphics[scale=0.5]{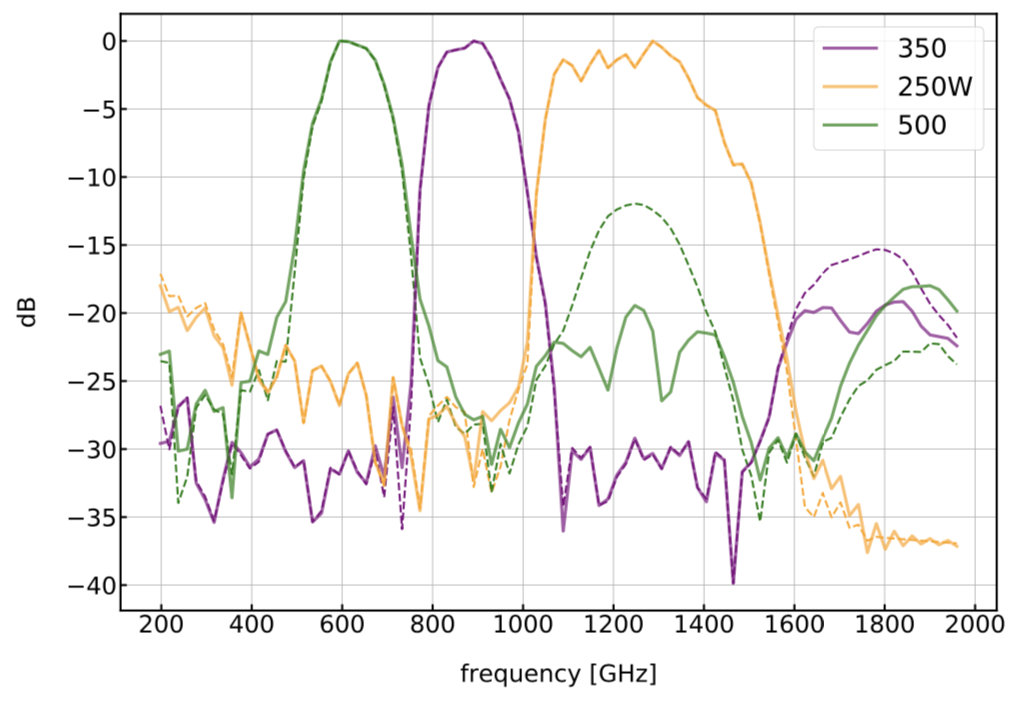}
    \caption{Bandpasses of the median channel for one 250 rhombus and the 350 and 500~$\mu m$ arrays from the measurements in Palestine.\cite{sam_thesis}}
    \label{fig:FTS}
\end{figure}

The last major new test on the receiver and detectors performed during the Palestine campaign was the measurement of the bandpasses for each of the arrays. This test was performed with a mercury arc lamp FTS which provided emission across spectrum of the instrument and an attached low-pass optical filter that blocked the high frequency radiation. The FTS aperture was mounted a few inches from the window of the receiver which allowed the source to fill the field of view of the receiver as well as minimized atmospheric absorption. To further ensure that each detector was well matched to the FTS the measurements were taken five times, with the FTS center, left, right, up, and down positioned relative to the window. The resulting inteferogram timestreams were processed to calculate the bandpass of each detector, and the median for each waveband can be seen in Figure \ref{fig:FTS}.

In order to optimize the overall noise performance of the system, one of the key considerations is that the power reaching the detectors must be at a certain level in order to read them out but not overdrive them. It is most desirable to attenuate the signal at the coldest stages as this will add the lowest ratio of thermal noise. By the end of the integration in Palestine we had managed to significantly improve our sensitivity estimates based on the measurements we had taken, but there was room for improvement by modifying the coaxial cable to reduce the warm attenuation and add additional attenuation at 1~K stage of the system. By doing this we shifted approximately 10~dB of attenuation on the input to the 1~K stage and by re-plumbing the output as well reduced the dependency on warm amplifier. The final measurements of detector sensitivity prior to Antarctic deployment are reported in Table \ref{tab:noisepal}.

\begin{table}[h]
    \centering
    \begin{tabular}{|c|c|c|}
    \hline
         Waveband & NEP Flight Planning & NEP Palestine Measurements\\ \hline
        250~$\mu m$ & $3.60*10^{-15}$ & $1.84*10^{-15}$ \\ \hline
        350~$\mu m$ & $2.83*10^{-15}$ & $1.84*10^{-15}$ \\ \hline
        500~$\mu m$ & $1.54*10^{-15}$ & $5.26*10^{-16}$ \\ \hline
    \end{tabular}
    \caption{Detector Noise-equivalent power (NEP) in $W \sqrt{Hz}^{-1}$. These are the optical NEP estimates and include planned (flight planning) and measured (measured receiver and planned mirror, Palestine) optical efficiencies of the instrument for direct use in comparison with astronomical sources. With the system upgrades we have notably improved the noise levels and thus mapping speed.}
    \label{tab:noisepal}
\end{table}

\subsection{Antarctica 2018}
\paragraph{}

Aside from a few mishaps that occurred during the shipping of the instrument to Antarctica, the campaign was relatively uneventful with respect to the readout and detectors. We repeated the sensitivity estimates and found that our numbers per detector had not changed from Palestine, as expected. Unfortunately, there were two components of the instrument that were damaged in shipping which significantly affected the capabilities of the instrument. The first was that the carbon fiber mirror was accidentally exposed to water en route to Antarctica which underwent a galvanic reaction and degraded the aluminum surface of the mirror, increasing emissivity and altering the beam within the degraded regions. Second, one of the RF output paths for the 250~$\mu$m array was damaged in shipping. This issue could not be detected until the receiver had been cooled down and by then it was too much of a time loss to fix and still fly that season. After many attempts to launch, the season was canceled due to the breakup of the circumpolar vortex.

\subsection{Antarctica 2019}
\paragraph{}
The BLAST team redeployed to Antarctica in 2019 having spent much of the free year working on system improvements. We developed a solution to fix the broken readout chain that also resulted in lower amplifier noise levels using a single amplifier with improved heat sinking. We also developed a vacuum-safe pump system which would lower the pressure in the helium pumped pot by a factor of 10, resulting in $>$20~mK gains in array temperature in flight. The mirror represented the last of the issues, and we experimented with several methods of mitigating the issue, such as mylar reflective tape and silver conductive paint. It was determined that the emissivity of the paint and microscopic features resulted in a net emissivity of $>$0.3. We instead opted for the cryogenic tape, with a reported value of $\sim 0.005$. Due to the higher difficulty in applying microstrips of tape we were only able to cover a fraction of the deteriorated surface.

\begin{wrapfigure}{R}{0.5\textwidth}
    \centering
    \includegraphics[scale=0.4]{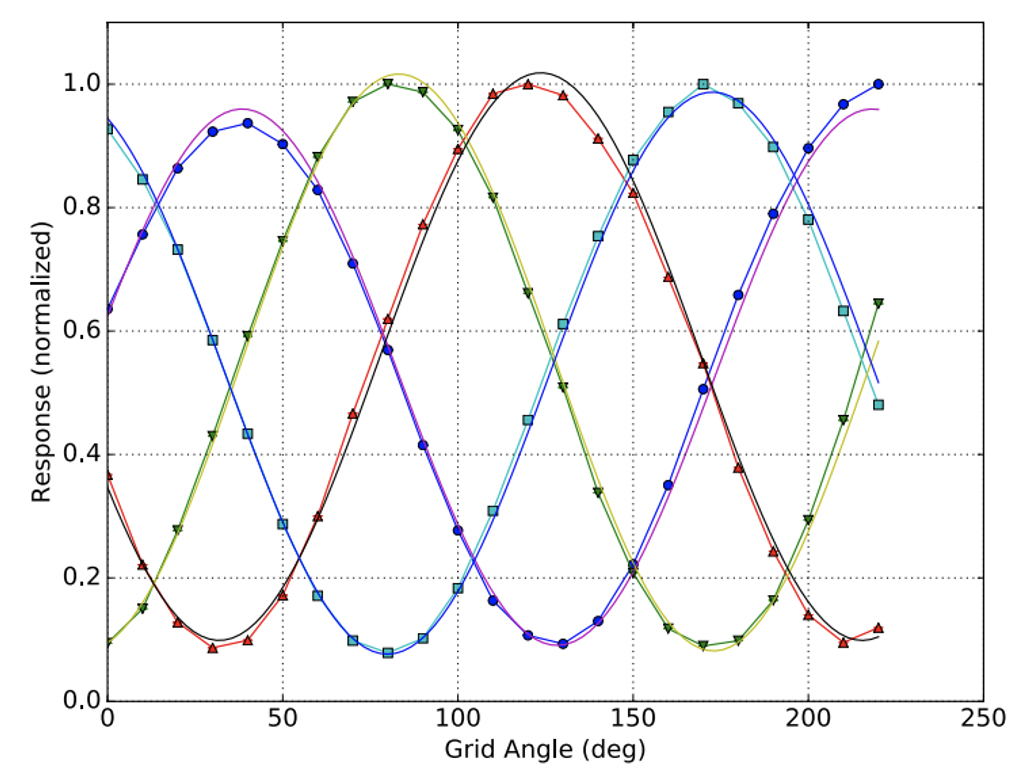}
    \caption{Measurements of the polarization response of the 4 different orientations of detectors for the BLAST-TNG 350~$\mu$m array.\cite{sam_ltd}}
    \label{fig:crosspol}
\end{wrapfigure}

Over the winter, the readout system had endured extremely cold temperatures, with the potential for component failure or malfunction. In addition, with the planned replacement of the broken RF pathway, the entire system would need to be recharacterized to ensure that the sensitivity remained the same. In light of this, we developed a new cryogenic beam filling source for noise measurements, an anti-reflective chopped IR source for array mapping, and a large chopped source for polarization measurements. Early testing indicated that the detectors and readout were working as expected which allowed us to proceed with the recharacterization.

During the campaign we performed the most detailed measurements of the polarization response of the detectors, showcasing the four orientations seen in Figure \ref{fig:crosspol}. The tests were performed with a polarizing grid filter which was rotated in increments of 10 degrees and the anti-reflective rotating chopper in front of a warmed AR plate. Time stream data of the chopped signals were taken over $\sim$30 seconds to ensure adequate on/off measurements before rotation. The resultant relative amplitudes were plotted relative to the rotation angle of the grid and fit to a sinusoidal model to determine the polarization efficiency. Results indicated that the polarization efficiencies were $>$0.86\cite{sam_ltd}. 

\begin{wrapfigure}{R}{0.4\textwidth}
    \centering
    \includegraphics[scale=0.6]{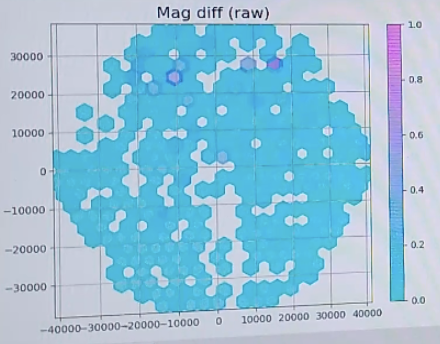}
    \caption{Live-plotted array map of the 350~$\mu$m array with several detectors responding to thermal noise in purple.}
    \label{fig:livemap}
\end{wrapfigure}

Another measurement we made was to remap the arrays using a new setup designed to minimize any reflective contamination in the system. In order to create a localized, choppable signal, we used an IR-56 parabolic focused source (Hawkeye Technologies) which was powered on and off at a frequency between 1 and 10Hz and scanned with the previously used stage. The source was surrounded by a grid of anti-reflective tiles\cite{xuMMAtile} to provide a constant background during the measurement. This apparatus was coupled to HDPE refocusing optics mounted to the front of the receiver which matched the detector beams to the source size. The source was then scanned across the entire field of view in the same manner as the previous measurement. The data were reduced and compared with the previous map from NIST, with good agreement. The spatial positions were used to create live-plotting software for the array which mapped detections to a color on a static array. An example of this software applied to the detectors can be seen in Figure \ref{fig:livemap}. These plots were used to ensure that all detectors were responding prior to further testing.

The final set of measurements we performed was the multiple static background noise measurements, similar to those described in Section \ref{sec:testing}. Because the different sources of noise in our system vary with the loading on the detectors, it was imperative to provide the coldest source with a known temperature that we could in order to elucidate the cold sky behavior. To this end, we designed a miniature cryostat with a window matched to the BLAST-TNG window which stared directly at the blackened surface of a liquid nitrogen tank. This provided a static 77K load, with which we could compare the system response to other static loads at warmer temperatures. These results allowed us to compare the predicted noise component values to those measured at different loadings and ensure confidence in the predicted in-flight values.
\section{In-flight Performance}
\label{section:inflight}
\paragraph{}
After a series of scrubbed attempts due to wind conditions, BLAST-TNG was launched from the NASA long-duration balloon facility just outside of McMurdo Station, Antarctica on January 6th, 2020. The flight was cut short due to the mechanical failure of a portion of the gondola after being struck by debris during the launch. Despite this fact, we were able to get valuable information about the functionality of many of the systems, as well as preliminary insights into the overall sensitivity of the instrument. In this section we discuss the in flight performance of the cryogenic receiver and the detectors.

\begin{wrapfigure}{R}{0.5\textwidth}
    \centering
    \includegraphics[scale=0.5]{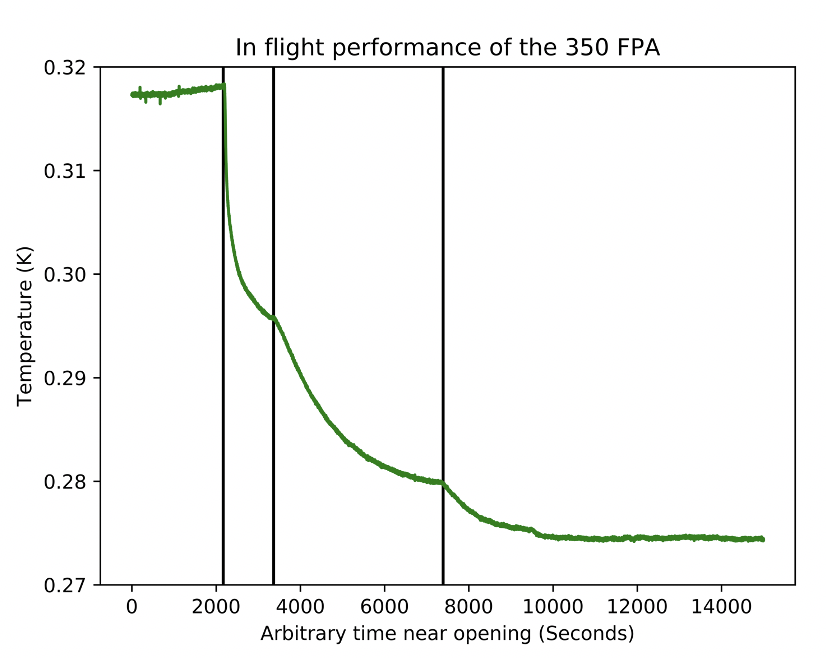}
    \caption{Focal-plane array temperatures for the 350~$\mu$m around the time the valves were opened to vent the pumped pot in-flight. Black vertical lines represent the opening to 6~mBar, 1.2~mBar, and 0.6~mBar pressures respectively.}
    \label{fig:opening}
\end{wrapfigure}

Over the course of the flight, the cryogenic receiver exceeded temperature expectations significantly. Previous in-lab testing had shown a base array temperature of 272-277~mK while the pumped pot was attached to a much larger scroll pump and 283-285~mK while attached to the flight system. Additionally, previous measurements in a 6~mBar test environment had indicated that our system could run at $>$300~mK, which would significantly degrade performance. In-flight, we performed a series of sequential openings in order to measure the equilibrium temperature of each configuration. The temperatures can be found in Table \ref{tab:cryotemps}, and a graph of the opening sequence can be seen in Figure \ref{fig:opening}. These in-flight temperatures represented an improvement over even the more powerful in-lab setup, and a marked gain in detector noise over the expectation.

\begin{table}[b]
    \centering
    \begin{tabular}{|c|c|c|c|}
    \hline
       Location  & Pot & Evaporator & Array\\ \hline
       0 pumps  & 1.56~K & 278~mK & 296~mK \\ \hline
       1 pump  & 1.28~K & 268~mK & 279~mK \\ \hline
       2 pumps  & 1.20~K & 265~mK & 274~mK \\ \hline
    \end{tabular}
    \caption{In flight equilibrium temperatures of the coldest stages of the cryogenic receiver. For the 0 pump case the pot is simply open to the atmosphere which is at approximately 6~mBar constant pressure.}
    \label{tab:cryotemps}
\end{table}

\begin{figure}[h]
    \centering
    \includegraphics[scale = 0.5]{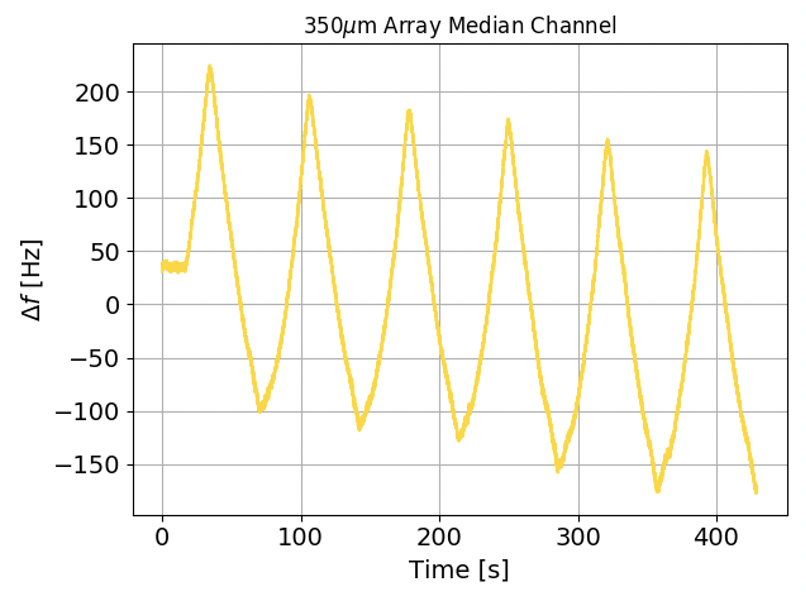}
    \includegraphics[scale = 0.5]{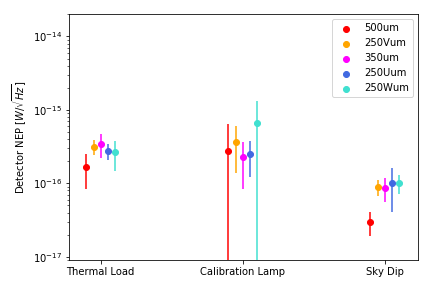}
    \caption{(left) Skydip response of the median 350~$\mu$m channel. (right) Sensitivities of the different BLAST-TNG readout chains calculated from static loads, the calibration lamp, and the skydips. Each has been scaled by the optical efficiency to provide an optical NEP.}
    \label{fig:sens_adrian}
\end{figure}

After initial system checkouts, the first observations we performed during the flight were a series of calibration measurements. The initial calibrator was to perform a series of "skydip" measurements where the telescope is held still in the azimuth direction and scanned quickly up and down in elevation. This provides a measurement of the response of the detectors to a changing load which is well-modeled by atmospheric simulation software. Additionally, this measurement is nearly independent of the beam and focus as the nominal beam size is under 1~\arcmin in all bands and we scan over an elevation range of more than 20~\degree. An example of several skydips in the timestream can be seen in Figure \ref{fig:sens_adrian} and the associated sensitivities are listed in Table \ref{tab:sense} and displayed compared to other measurements in Figure \ref{fig:sens_adrian}. While this test gives an indication of the noise, it does not provide any description of the beam or point source sensitivity. Both of these required us to scan and map a small source which had been previously measured by another instrument for direct comparison. Due to the visibility constraints, our first choice was unavailable for the majority of the flight and our initial calibration scans were of the source RCW92B. Efforts to reduce the data from these scans are ongoing while all data are reduced from critical subsystems.

\begin{table}[t]
    \centering
    \begin{tabular}{|c|c|c|c|}
    \hline
         &  250~$\mu m$&350~$\mu m$&500~$\mu m$\\
         \hline
        $NEP_{FLIGHT}$ & $1.48*10^{-15}$ &$6.90*10^{-16}$&$3.61*10^{-16}$\\
        \hline
    \end{tabular}
    \caption{Noise levels calculated from the skydips performed during the BLAST-TNG flight. The change in incident power for a skydip is calculated using the AM atmospheric model software\cite{paine_atmospheric} which estimates loading from the sky for given coordinates and times.}
    \label{tab:sense}
\end{table}

\section{Conclusions}
\label{section:concs}
\paragraph{}
The BLAST-TNG experiment is a groundbreaking instrument designed to showcase the viability of kilopixel MKID arrays in the stratosphere. We started from a bare-bones readout system using either a slow-sweeping VNA to map the resonances or watching a single resonance in the time domain. From there we built up the ROACH2 system to map and then readout up to 1000 detectors simultaneously over a single coaxial line. We also worked within the confines of a sorption fridge cooling system to hit the goal array temperatures of 275~mK or lower. Through testing and implementation of the valved pot and scroll pump systems we were able to significantly reduce the parasitic heat load on the arrays, resulting in the 274~mK temperatures seen in flight. By the time we flew, software developed for our instrument had already been deployed on OLIMPO and tested for TolTEC. While our flight was cut short, we were able to demonstrate excellent noise properties in the detectors and their resilience to stratospheric concerns. With the knowledge and experience gained by developing these systems for the BLAST-TNG payload, future MKID experiments will be able to utilize them with relative ease.

\bibliography{References}
\bibliographystyle{spiebib} 

\end{document}